\documentclass[vecphys]{svmult}

\usepackage{makeidx}         % allows index generation
\usepackage{graphicx}        % standard LaTeX graphics tool
\usepackage{multicol}        % used for the two-column index
\usepackage[bottom]{footmisc}% places footnotes at page bottom
\makeindex             % used for the subject index
\begin{document}

\title*{Evolutionary synthesis models for spirals and
irregular galaxies}
\titlerunning{Synthesis models for spiral galaxies} 
%for an abbreviated version of
\author{Mercedes Moll\'{a}}

\institute{CIEMAT, Avda. Complutense 22, 28040, Madrid
\texttt{mercedes.molla@ciemat.es}}
%\and Name and Address of your Institute \texttt{name@email.address}}
\maketitle
\index{Mercedes Moll\'{a}}
%\index{Author2}
% Use the \index{} command to code your author index

\abstract{
We show autoconsistent chemical and spectro-photometric evolution
models applied to spiral and irregular galaxies.  Evolutionary
synthesis models usually used to explain the stellar component
spectro-photometric data, are combined with chemical evolution models,
to determine precisely the evolutionary history of spiral and
irregular galaxies. In this piece of work we will show the results
obtained for a wide grid of modeled theoretical galaxies.}

\section{Introduction}
\label{intro}

It is well known that galaxies have different spectral energy
distributions (SEDs) depending on their morphological type
\cite{col80}.  These SEDs, and other data related to the stellar
phase, are usually analysed through (evolutionary) synthesis models 
(see \cite{rosa05,mar05} for a recent and updated review about these
models), based on single stellar populations (SSPs), created by an
instantaneous burst of star formation (SF).  These codes compute the
SED, $S_{\lambda}$, and the corresponding colors, surface brightness
and/or spectral absorption indices emitted by a SSP of metallicity $Z$
and age $\tau$, from the sum of spectra of all stars created and
distributed along a HR diagram, convolved with an initial mass
function. This SED, given $\tau$ and Z, is characteristic of each SSP.

SF, however, does not always take place in a single burst, as occurs
in spiral and irregular galaxies where star formation is continuous or
in successive bursts. In that case, the characteristic $\tau$ and Z
found after fitting a SSP model to an observed SED represent only
averaged values, leading to a loss of information. Our aim is to
obtain the time evolution of the galaxy, given by the SF history (SFH)
and the age-metallicity relation (AMR).  When more than a single
generation of stars exists in a region or galaxy, the final SED, $\rm
F_{\lambda}$, corresponds to the light emitted by successive
generations of stars. It may be calculated as the sum of several SSP
SEDs, $S_{\lambda}$, being weighthed by the created stellar mass in
each time step,implying a convolution with the SFH, $\Psi(t)$:

\begin{equation}
F_{\lambda}(t)=\int_{0}^{t} S_{\lambda}(\tau,Z)\Psi(t')dt',
\label{Flujo}
\end{equation}
where $\tau=t-t'$.

This requires to us make hypotheses about the shape and the intensity
of the SFH, e.g. an exponentially decreasing function of time is
usually assumed.  The fit will give the best parameters that are
needed to define the assumed function. In fact, in order to avoid
bias, one would need to consider also other possible SFHs and their
parameters fits, but computational resources precluded the study of
such large model grids.  Moreover. an important point, usually
forgotten when this technique is applied, is that
$S_{\lambda}(\tau,Z)= S_{\lambda}(\tau,Z(t'))$, that is, the
metallicity changes with time since stars form and die continously.
It is not clear which $Z$ must be selected at each time step without
knowing this function $Z(t)$. Usually, only one $Z$ is used for the
whole integration which may be an over-simplification.

A better method (\cite{cid05}) is to perform a least square technique
to find the best superposition of these SSP's which will fit the
data. This gives the proportions of stellar mass created at a given
metallicity and age that better reproduce observations, thus
providing the SFH, $\Psi(t)$, and the AMR, $Z(t)$, as results.  This
technique, although essentially correct, produces, at the moment, poor
evolutionary histories, having low time resolution due to 
large bin widths.

In summary, it is clear that spiral and irregular galaxies are systems
more complex than those represented by SSP's, and that, in particular,
their chemical evolution must be taken into account for a precise
interpretation of the spectro-photometric data. On the plus side for
these objects the gas phase data are also available and may be used as
constraint.  What is required then is to determine the possible
evolutionary paths followed by a galaxy that arrive at the observed
present state, while, simultaneously, reproducing the average
photometric properties defined by the possible subjacent stellar
populations.

\section{Chemical evolution models}
\label{models}

\begin{figure}[b]
\centerline{%
\begin{tabular}{c@{\hspace{2pc}}c}
\includegraphics[width=0.4\textwidth,angle=-90]{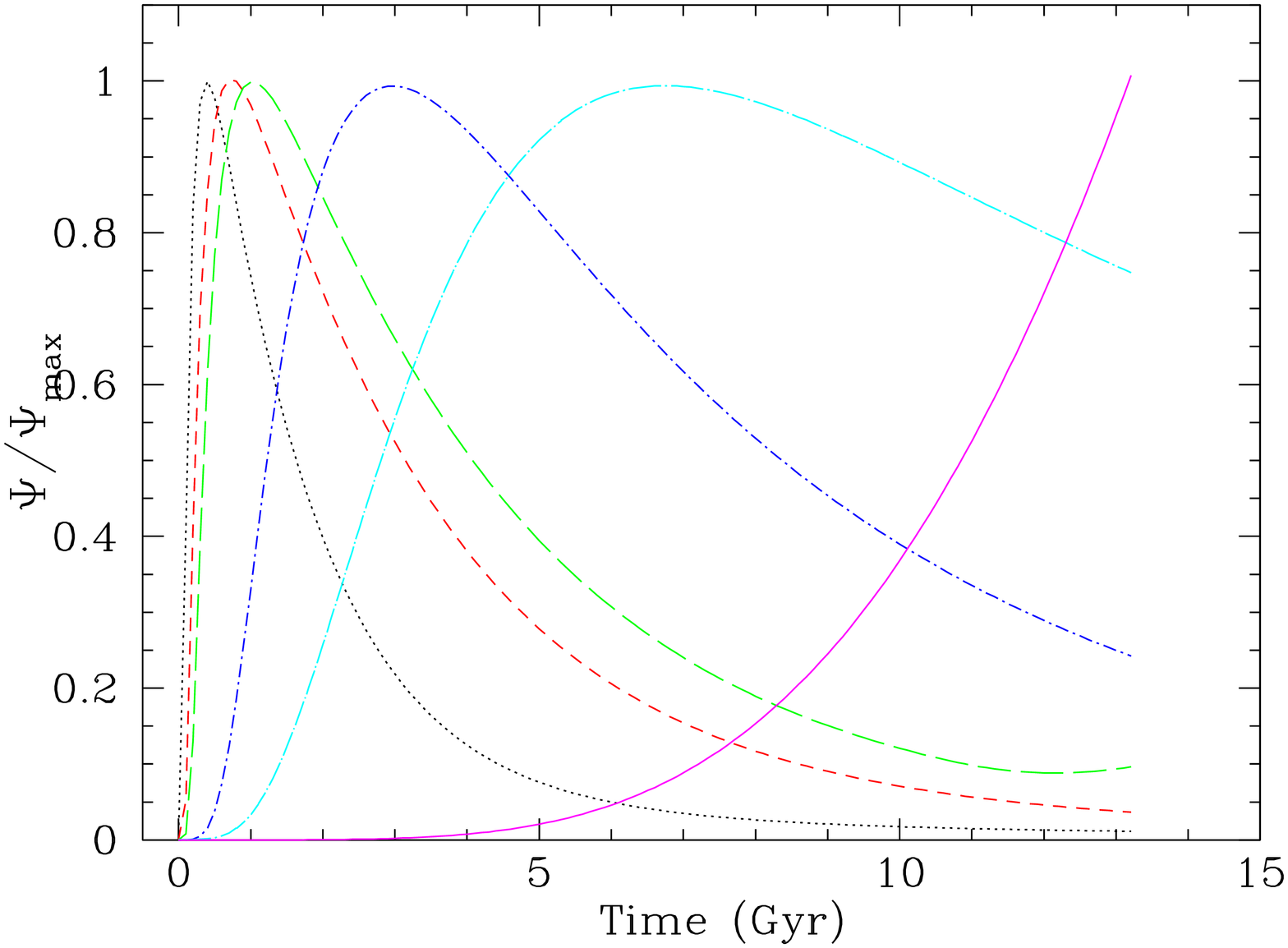} &
\includegraphics[width=0.4\textwidth,angle=-90]{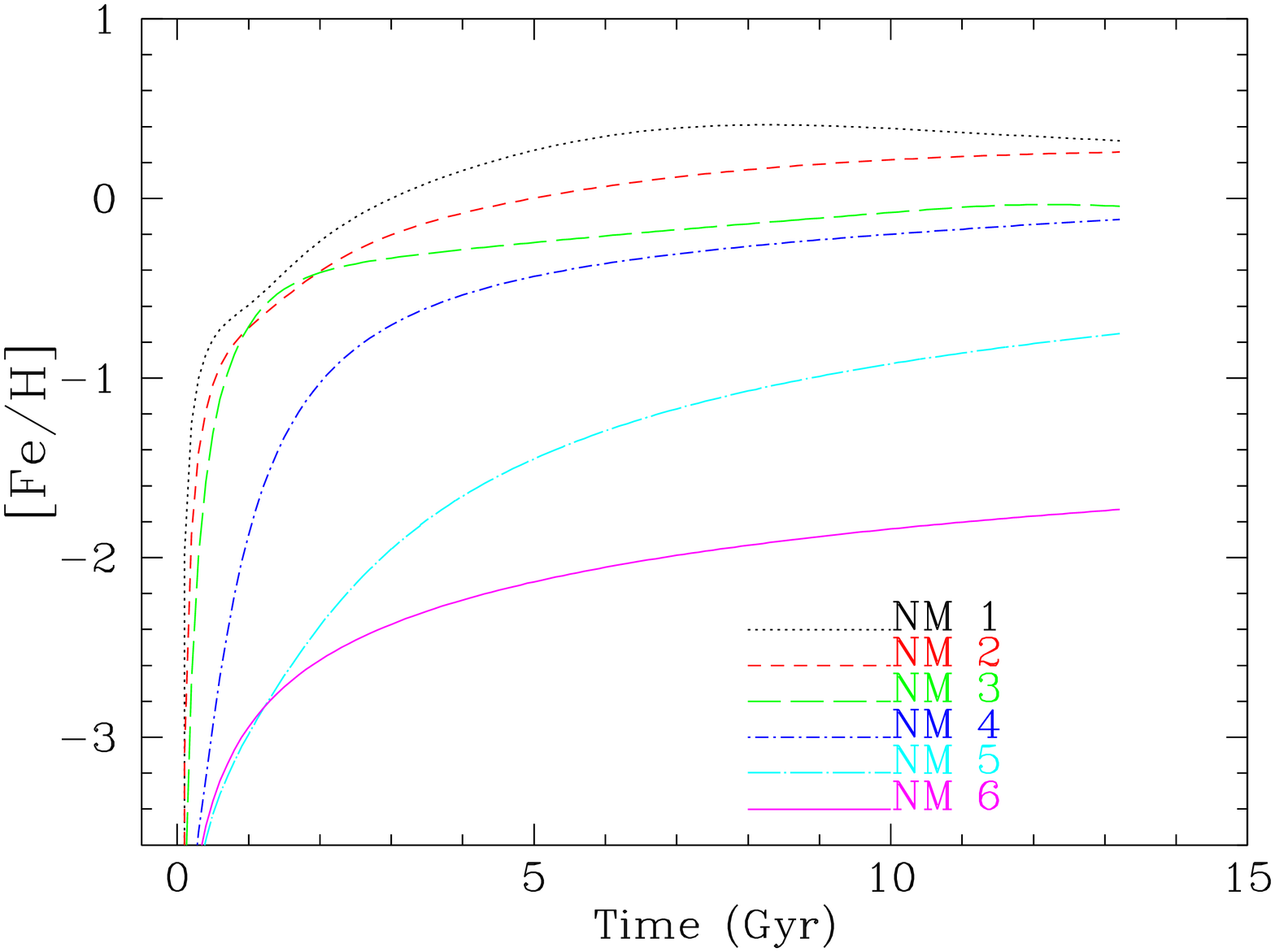}\\
\end{tabular}}
\caption[]{The resulting SFH, $\Psi(t)$, and  AMR, [Fe/H](t) for models
of Table 1}
\label{sfh}       
\end{figure}
 
The information coming from the gas phase, such as density, abundance,
and actual star formation rate, is usually analysed using chemical
evolution models.  They describe how the proportion of heavy elements
present in the interstellar medium (ISM) increases, starting from
primordial abundances, when stars evolve and die.  Modern codes solve
numerically the system of equations used to describe a scenario based on
initial conditions for the total mass of the region, the existence of
infall or outflow of gas, and the initial mass function (IMF).
Stellar mean-lifetimes and yields, known from stellar evolutionary
tracks are also included.  Finally, a SF law is assumed.
The {\sl best} model for a galaxy will of course be the one which
reproduces the observational data as closely as possible.

\begin{table}
\centering
\caption{Selected models similar to some known galaxies}
\label{table1}       
\begin{tabular}{llllllllcrl}
\hline\noalign{\smallskip}
Ndis & Vrot & N & $\rm R_{opt}$ & $\tau_{col}$ & $\Psi_{max}$ & $t_{\Psi,max}$  & [Fe/H]$_{p}$ &Galaxy &T & Type\\
     & km.s$^{-1}$ &  & kpc& Gyr & $\rm M_{\odot}.yr^{-1}$ &  Gyr &  \\
\noalign{\smallskip}\hline\noalign{\smallskip}
   43&  371 & 1  &      29  &    1.65 &  318   &   0.32 & 0.322 & M81 & 2 & Sab  \\
   38&  266 & 3  &      21  &    2.66 &  78    &   0.60 & 0.258 & M31 & 3 & Sb \\
   28&  200 & 4  &      13  &    4.00 &  32.39 &   1.05 & -0.04 & MWG & 4 & Sbc\\
   18&   99 & 6  &       7  &    11.00&  2.78  &   3.02 & -0.12 & M33 & 6 & Scd\\
   13&   87 & 8  &       4  &    13.3 &  0.20  &   6.68 & -0.75 & NGC300 & 7 & Sdm\\
   8 &   78 & 10 &      2.3 &    17.7 &  0.0013&   13.2 & -1.73 & D0154 & 10 & Im\\
\noalign{\smallskip}\hline
\end{tabular}
\end{table}

We use here the multiphase chemical evolution model, described
in \cite{fer92,fer94,mol96}. We have computed in \cite{mol05} a large
grid of models for 44 theoretical galaxies with variable initial
masses and sizes obtained with the universal rotation curve of
\cite{pss96}.  The mass in each modeled galaxy, initially in the form of
gas, is in a protohalo, from which it falls onto an equatorial plane
leading to the formation of the disc. The gas infall rate, or its inverse the
collapse-time scale, depends on the total mass.  For each of these
44 galaxies we assumed 10 possible molecular cloud and SF
efficiencies, $\epsilon$, which may take values in the range [0,1] and
which are distinguished by the number N. In this way we computed models
in the whole plane $\rm [Vrot, N]$ (or Vrot-$\epsilon$).  Within this
plane, the bright galaxies are located on the diagonal axis,
where the most massive spiral galaxies are usually those of earliest
morphological types \cite{rob94} which, in turn, need the highest star
formation efficiencies (see Fig.4 from \cite{mol05}).  Furthermore,
there exist other well defined zones in this plane 
corresponding to other classes of galaxies, such as the low mass dwarf
galaxies ($\rm Vrot< 70$ km.s$^{-1}$).

This grid reproduces well the generic data trends for the normal
bright spiral and irregular galaxies, in particular the radial
distributions observed for the gas, the stars, the SFR
and the elemental abundances in particular galaxies (see \cite{mol05}
for details). Each model produces the time evolution within a galaxy,
that is, the resulting SFH and AMR for several radial regions.  We
show these evolutionary histories in Fig.~\ref{sfh} for the selected
models given in Table~1. In this table we give the number of the mass
distribution $Ndis$ and its corresponding rotation velocity $Vrot$ in
columns 1 and 2, the number N which defines the SF
efficiency in column 3, and the disk radius $\rm Ropt$ and collapse
time scale $\tau_{col}$ in columns 4 and 5. The SF, as we
may see in panel a) of Fig.\ref{sfh} has a maximum $\Psi_{max}$ ( with
which we have normalized the curves of panel a)), at a given time
$t_{\Psi,max}$ which is different for each model. Both values are
given in columns 6 and 7 of the same table. In column 8 we give the
present time Iron abundance $\rm [Fe/H]_{p}$. Each one of these models
is representative of a well known galaxy as those referenced in column 9,
as checked in \cite{mol05}.

\section{Evolutionary synthesis models}

\begin{figure}[b]
\centering
\includegraphics[width=0.7\textwidth,angle=-90]{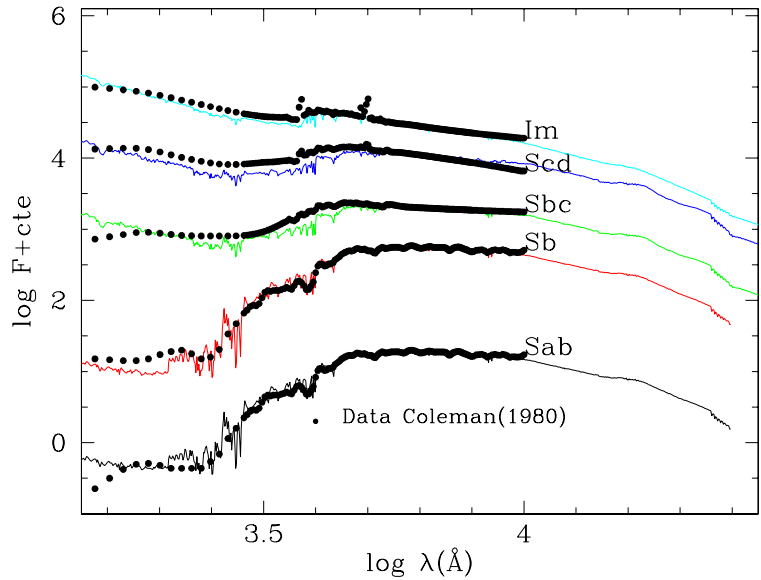} 
\caption[]{The resulting spectral energy distributions obtained
for models of Table~1 compared with the fiducial templates from \cite{col80}.}
\label{sp_obs}
\end{figure}

From the results of the above section, we take the SFH and the AMR as inputs
in Eq.~\ref{Flujo} to compute the SED of each galaxy.  The set of
SSP's SEDs used are those from the evolutionary synthesis code
described in \cite{mol00}.  For each stellar generation created in the
time step $t'$, a SSP-SED, $S(\tau)$, from this set is chosen taking
into account its age, $\tau=t-t'$, from the time $t'$ in which it was
created until the present $t$, and the metallicity $Z(t')$ reached by
the gas. After convolution with the SFH, $\Psi(t)$, the final SED,
$F(\lambda)$, is obtained.

The resulting $\rm F(\lambda)$ reproduce reasonably well the SEDs of
galaxies such as can be seen in Fig.~\ref{sp_obs}.  There we compare
our resulting spectra for the models of the known galaxies of
Table~\ref{table1} to the different morphological type templates from
\cite{col80}.

In order to use our model grid, we may therefore to select the best
model able to fit a given observed SED and then see if the
corresponding SFH and the AMR of this model are also able to reproduce
the present time observational data of SFR and
metallicity of the galaxy.

\begin{figure}
\centering
\includegraphics[width=0.8\textwidth,angle=0]{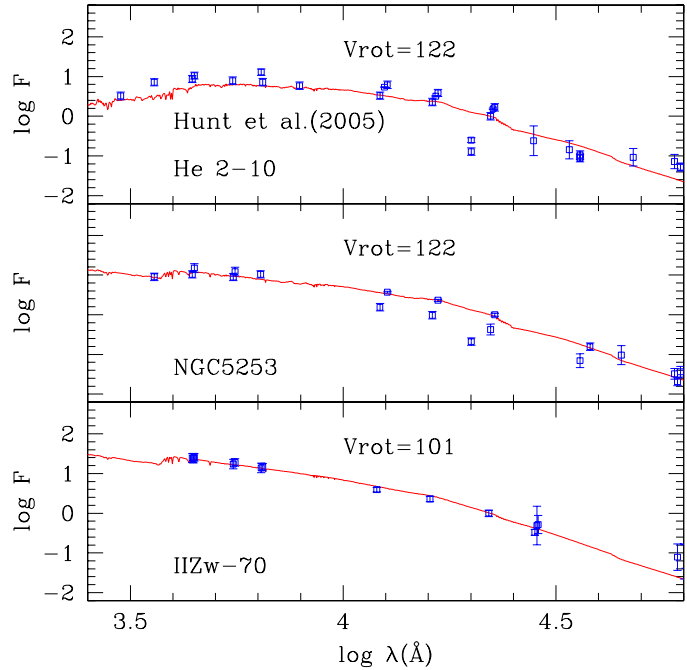} 
\caption{The resulting spectral energy distributions --red solid
lines- obtained to reproduce the observations from \cite{hunt06}
--blue open squares- for three BCD galaxies.}
\label{sp_bcd}
\end{figure}

\begin{figure}
\centering
\includegraphics[width=0.8\textwidth,angle=0]{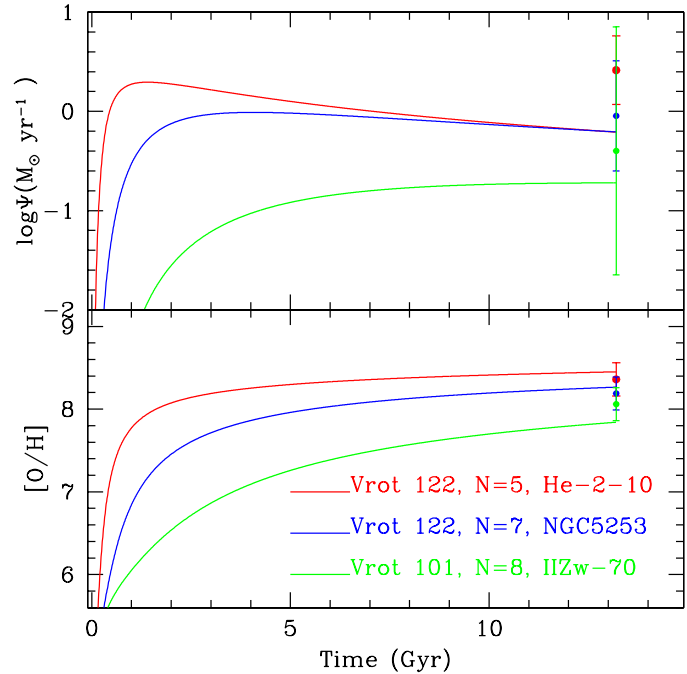}
\caption{The resulting SFH and AMR able to fit the SEDs of
Fig.~\ref{sp_bcd} represented by the blue, red and green lines for
He-2-10, NGC5253 and IIZW-70, respectively.  The present time data are
shown in both panels by the full dots with error bars.}
\label{sfh_bcd}
\end{figure}

We show an example of this method in Fig.~\ref{sp_bcd} where some SEDs
from \cite{hunt06} of BCD galaxies are compared with the best model
chosen for each one of them. In Fig.~\ref{sfh_bcd} we show the
corresponding SFH and AMR with which the modeled SEDs were
computed. The final values are within the error bars of observations
for these same galaxies, compiled by the same authors \cite{hunt06}.
Since each SED is well fitted and, simultaneously, the corresponding
present--time data of the galaxy by the same chemical evolution
model, we may be confident that these SFH and AMR give to us a reliable
characterization of the evolutionary history of each galaxy.

\section{Conclusions}

We have computed a grid of 440 chemical evolution models for spiral
and irregular galaxies of different sizes and masses ($\rm
Vrot=40$--400 km.s$^{-1}$).

For each one of them we assumed 10 different SF efficiencies and the
corresponding time evolutionary histories as given by the SFH and the
AMR, were then used as inputs to obtain the SED through Eq.(1).

In this way we have obtained spectral energy distributions, colors,
surface brightness and absorption spectral indices for each
theoretical galaxy of the grid.

This combination of techniques allows the use of two data types: the
ones proceeding from the gas phase as well as that coming from
spectro-photometry.  If a model is able to reproduce simultaneously
both sets of observations for a given galaxy, the SFH and the AMR
resulting from the corresponding chemical evolution model may be
considered sufficiently reliable to represent its very evolutionary
history.

%
%
% BibTeX users please use
%\bibliographystyle{/ae37a/mercedes/papers/apj.bst}
%\bibliography{bibliografia}
%
% Non-BibTeX users please follow the syntax
% the syntax of "referenc.tex" for your own citations
%\input{referenc}
%%%%%%%%%%%%%%%%%%%%%%%%%%%%%%%%%%%%%%%%%%%%%%%%%%%%%%%%%%%%%%%%%%%%%%  }

%%%%%%%%%%%%%%%%%%%%%%%%%%%%%%%%%%%%%%%%%%%%%%%%%%%%%%%%%%%%%%%%%%%%%%

\printindex
\end{document}